\documentclass[conference]{IEEEtran}
\IEEEoverridecommandlockouts
\usepackage{bbm}
\usepackage[font=small]{caption}
\usepackage{cite}
\usepackage{amsthm}
\usepackage{amsmath,amssymb,amsfonts}
\usepackage{algorithmic}
\usepackage{graphicx}
\usepackage{textcomp}
\usepackage{xcolor}
\usepackage{booktabs}
\usepackage{tabularx}
\usepackage{graphicx}
\usepackage{comment}
\usepackage[ruled,vlined]{algorithm2e}
\usepackage{subcaption}  
\usepackage{subfig}

\def\BibTeX{{\rm B\kern-.05em{\sc i\kern-.025em b}\kern-.08em
    T\kern-.1667em\lower.7ex\hbox{E}\kern-.125emX}}

\begin{document}

\title{CA3D: Computing Accessibility-Aware Cooperative 3D Deployment of Multiple UAVs\\  
\thanks{The work of Yiqin Deng was supported in part by the National Natural Science Foundation of China (Grant No. 62301300) and by the Shandong Provincial Natural Science Foundation (Grant No. ZR2023QF053). The work of Yuguang Fang was supported in part by a grant from the Research Grants Council of the Hong Kong Special Administrative Region, China (Project No. CityU 11216324). }
\thanks{Yiqin Deng is with School of Data Science, Lingnan University, Tuen Mun, Hong Kong, China (email: yiqindeng@ln.edu.hk). Zihan Fang, Qingxiao Huang, Junhui Gao, Qianyao Ren, and Yuguang Fang are with Hong Kong JC STEM Lab of Smart City and Department of Computer Science, City University of Hong Kong, Kowloon, Hong Kong, China (email: zihanfang3-c@my.cityu.edu.hk, qx.huang@my.cityu.edu.hk, junhui.gao@cityu.edu.hk, qianyao.ren@my.cityu.edu.hk, my.fang@cityu.edu.hk). Yijie Wang is with Institute of Big Data, Central South University, Changsha, China (email: wangyijiecs@csu.edu.cn).
}}

\author{\IEEEauthorblockN{Yiqin Deng, Zihan Fang, Yijie Wang, Qingxiao Huang, Junhui Gao, Qianyao Ren, and Yuguang Fang
}
}

\maketitle
\thispagestyle{plain}


\begin{abstract}
This letter investigates computing-accessibility-aware cooperative 3D deployment of multiple UAVs for task completion enhancement, termed \textit{CA3D}. We first provide a theoretical analysis showing that computing accessibility is the key mechanism linking UAV deployment to delay-constrained task completion, and that UAV inter-spacing creates a fundamental tradeoff between computing-resource accessibility and task completion. We then develop a cooperative 3D deployment design that jointly balances accessible computing capacity, task completion probability, and redundant UAV overlap. Simulation results under heterogeneous computing node capacities show that \textit{CA3D} consistently outperforms \textit{Random}, \textit{Fixed}, and \textit{Greedy} deployment baselines under both hotspot and random ground user (GU) distributions. Under the hotspot GU distribution, \textit{CA3D} achieves nearly full task completion, improving the task completion probability by about \(3.3\times\) over \textit{Random} deployment when the number of UAVs is \(8\). Under a more challenging random GU distribution, \textit{CA3D} still achieves about \(35\%\) higher task completion probability than the best baseline when the number of UAVs is \(12\). These results demonstrate that computing-accessibility-aware cooperative 3D deployment improves not only task completion but also robustness to GU distribution changes.
\end{abstract}


\section{Introduction}
\label{sec:intro}
Unmanned aerial vehicles (UAVs) have emerged as a flexible platform for extending wireless communication and edge computing services in infrastructure-limited environments~\cite{Deng2024,Deng_2026,Deng2024twc}.
In particular, UAV-enabled ground computing networks, such as UAV-assisted multi-access edge computing (MEC)~\cite{Han2024}, have attracted significant attention because they can provide on-demand connectivity and task offloading opportunities for ground users when terrestrial infrastructure is unavailable, congested, or temporarily disrupted. 
By adjusting their positions in three-dimensional (3D) space, UAVs can improve air--ground link quality, enlarge service coverage, and create more favorable conditions for delay-sensitive task offloading to ground computing nodes. Compared with terrestrial gateways, UAV gateways provide unique
advantages in 3D mobility, rapid on-demand deployment, and elevated
line-of-sight connectivity. They can be dynamically repositioned between local user hotspots and geographically distributed CNs, making them particularly suitable for bridging local
task demand to geographically distributed computing resources when terrestrial gateways are unavailable, congested, or spatially mismatched with the task demand.
Motivated by these advantages, recent studies have investigated multi-UAV deployment, trajectory optimization, and user association to improve communication and computation performance in UAV-enabled ground computing networks~\cite{Zeng20243D, Zhu2024multi,Han2024}.

Despite these advances, most existing multi-UAV MEC frameworks assume that candidate computing nodes are prefixed and spatially bounded, typically corresponding to edge servers deployed within the coverage area of users for task computing. Under this paradigm, UAV deployment primarily serves to improve communication links between users and a fixed set of computing nodes. However, this assumption becomes restrictive in emerging computing power networks (CPNs), where heterogeneous computing resources are geographically distributed across large areas, including roadside units, micro data centers, and vehicular computing platforms~\cite{Chen2024}. In such networks, UAVs act as mobile gateways that dynamically bridge users to dispersed computing resources. As a result, the set of ground computing nodes that can be accessed is not fixed, but instead depends on the UAV deployment and the end-to-end latency constraints. In principle, UAVs can potentially connect users to a much larger pool of computing resources beyond their co-locating immediate coverage area~\cite{Deng2024,Deng_2026}.

This observation leads to a fundamentally different perspective for UAV-enabled computing power networks (UAV-CPN). 
In such networks, UAV deployment does not merely improve communication quality, but also reshapes the network-wide set of computing nodes that can be effectively reached under latency constraints~\cite{deng2025uav_c}. 
In particular, the 3D placement of multiple UAVs determines how much geographically distributed computing resources can be made accessible to ground users, while UAV cooperation can expand the unique accessible computing pool and reduce redundant overlap among UAV service regions. 
As a result, computing accessibility emerges as a key enabler linking multi-UAV deployment to delay-constrained task completion~\cite{Deng2024,Deng_2026}.

Motivated by this insight, this letter investigates computing-accessibility-aware cooperative 3D deployment of multiple UAVs for task completion enhancement. 
We first provide a theoretical analysis showing that UAV inter-spacing offers a fundamental tradeoff between enlarging the unique accessible computing capability and maintaining favorable computing service conditions. 
Guided by this analysis, we develop a cooperative 3D deployment design that jointly balances computing accessibility, communication link quality, and task completion. 
Simulation results show that the proposed framework significantly improves both accessible computing capability and task success probability over representative baselines, demonstrating the value of cooperative multi-UAV deployment in transforming spatially distributed computing resources into effective service gains.

\section{System Model}

Consider a multi-UAV-enabled CPN, consisting of a set of ground users (GUs) $\mathcal{K}=\{1,\ldots,K\}$, a set of ground computing nodes (CNs) $\mathcal{N}=\{1,\ldots,N\}$, and a set of UAVs $\mathcal{M}=\{1,\ldots,M\}$. The ground CNs represent heterogeneous distributed computing resources,
such as roadside edge servers, micro data centers, and vehicular computing
platforms. Different CNs may have different available computing capabilities
due to their hardware configurations, service types, and background workloads.
We denote by \(f_n\) the available computing capability of CN \(n\), and define
\(\mathbf f=[f_1,\ldots,f_N]\) as the CN-capacity vector. In general,
\(f_n\) is node-dependent and is not assumed to be identical across CNs. Unlike conventional multi-UAV-assisted MEC systems, where candidate computing nodes are prefixed within a bounded service area, the CNs considered here are geographically distributed over a much larger region and may be located much farther away. Consequently, their accessibility depends critically on the UAV deployment.

Let the 3D position of UAV $m\in\mathcal{M}$ be denoted by $\mathbf{q}_m=[x_m,y_m,h_m]$, where $(x_m,y_m)$ and $h_m$ are its horizontal coordinates and altitude, respectively. Each UAV serves as a mobile gateway that collects tasks from nearby users and forwards them to feasible ground CNs. The task generated by user $k$ is characterized by $\Gamma_k=(L_k,C_k,D_k^{\max})$, where $L_k$, $C_k$, and $D_k^{\max}$ denote the input data size, required CPU cycles, and the maximum tolerable end-to-end latency, respectively.

For UAV $m\in\mathcal{M}$ located at $\mathbf{q}_m=[x_m,y_m,h_m]$, let $r_{k,m}$ denote the horizontal distance between user $k$ and UAV $m$, and let $r_{m,n}$ denote the horizontal distance between UAV $m$ and CN $n$. The corresponding link distances are $d_{k,m}=\sqrt{r_{k,m}^2+h_m^2}$ and $d_{m,n}=\sqrt{r_{m,n}^2+h_m^2}$.

Following the probabilistic air-ground channel model in~\cite{Khuwaja2018}, the line-of-sight probability for the user--UAV link is
$P_{k,m}^{\rm LoS}={1}/{1+a\exp\!\left(-b(\theta_{k,m}-a)\right)}$,
where $a$ and $b$ are environment-dependent constants and 
$\theta_{k,m}={180}/{\pi}\arctan\!\left({h_m}/{r_{k,m}}\right)$ is the elevation angle. The average large-scale channel gain is 
$g_{k,m}=\beta_0 d_{k,m}^{-\alpha_u}\left[P_{k,m}^{\rm LoS}+\eta(1-P_{k,m}^{\rm LoS})\right]$,
where $\beta_0$ is the reference channel gain, $\alpha_u$ is the path-loss exponent, and $\eta\in(0,1)$ denotes the additional attenuation for non-line-of-sight links. For simplicity of analysis, we assume orthogonalized access/backhaul spectrum resources among concurrent UAV-user and UAV-CN transmissions, so that inter-UAV interference is negligible. This allows us to isolate the impact of cooperative 3D deployment on computing accessibility and delay-constrained task success.
The achievable uplink transmission rate from user $k$ to UAV $m$ is 
$R_{k,m}=B\log_2\!\left(1+{p_k g_{k,m}}/{\sigma^2}\right)$,
where $B$ is the bandwidth, $p_k$ is the user transmit power, and $\sigma^2$ is the noise power. The corresponding uplink delay is $D_{k,m}^{\rm up}={L_k}/{R_{k,m}}$.

Similarly, the UAV--CN forwarding link follows the same propagation model, where the channel gain is denoted by $g_{m,n}$ with distance $d_{m,n}$. The achievable transmission rate is $R_{m,n}=B\log_2\!\left(1+{p_m g_{m,n}}/{\sigma^2}\right)$, with $p_m$ being the transmit power of UAV $m$. The corresponding forwarding delay for task $k$ is $D_{m,n}^{\rm fwd}={L_k}/{R_{m,n}}$.

If the task of GU \(k\) is executed at CN \(n\), the computation delay is
approximated as $D_{k,n}^{\mathrm{cmp}} = C_k / f_n$.
This node-dependent parameter captures the heterogeneity among different
types of CNs. For example, a micro data center may provide a larger \(f_n\)
than a vehicular computing platform, while the available value can also vary
with the current background workload. Since the result size is typically much smaller than the task input, the return delay is neglected as in~\cite{Deng_2026}. The end-to-end latency via UAV \(m\) and CN \(n\) is therefore
$D_{k,m,n}=D_{k,m}^{\rm up}+D_{m,n}^{\rm fwd}+D_{k,n}^{\rm cmp}$. Note that the UAV 3D position $\mathbf{q}_m=[x_m,y_m,h_m]$ affects both the link distances and elevation angles, thereby influencing the achievable transmission rates and the end-to-end latency.

For UAV \(m\), we define its \emph{accessible computing set} as
$\mathcal{C}_m(\mathbf{q}_m)
=
\{\, n\in\mathcal{N} \mid \exists k:\; D_{k,m,n} \le D_k^{\max} \,\}$.
That is, a ground computing node \(n\) belongs to
\(\mathcal{C}_m(\mathbf{q}_m)\) if there exists at least one user task that
can be completed through UAV \(m\) and CN \(n\) within its end-to-end latency
requirement. Therefore, \(\mathcal{C}_m(\mathbf{q}_m)\) represents the set of
ground CNs that are feasible for UAV \(m\) under the current 3D deployment.
Since \(D_{k,m,n}\) depends on the user--UAV transmission delay, the UAV--CN
forwarding delay, and the computation delay \(C_k/f_n\), this set is jointly
shaped by the UAV position and the node-dependent computing capability \(f_n\).
A CN with a larger \(f_n\) may become accessible even when its forwarding link
is relatively long, while a low-capacity CN may be infeasible under the same
latency deadline. In contrast to conventional MEC, the candidate CNs are not
fixed beforehand, but are deployment-dependent and may vary with the UAV's
horizontal location, altitude, and the heterogeneous CN capacities. Here,
accessibility is defined from the UAV perspective: a CN is considered
accessible to UAV \(m\) if it can support at least one user task served by UAV
\(m\) within the corresponding latency deadline.


To characterize the deployment benefit, we define the \emph{unique accessible computing capacity} as
$$\Psi(\mathbf{Q}) = \sum_{n \in \bigcup_{m \in \mathcal{M}} \mathcal{C}_m(\mathbf{q}_m)} f_n,$$ 
where $\mathbf{Q}=\{\mathbf{q}_m,\,m\in\mathcal{M}\}$ denotes the multi-UAV deployment. This metric captures the total non-overlapping computing capacity that becomes accessible under the current UAV deployment. To better understand how UAV spatial separation affects deployment-shaped computing accessibility, we next analyze a simplified two-UAV special case.

\section{A Two-UAV Case Analysis}

To gain analytical insight into the impact of UAV spatial deployment on computing accessibility, we consider a simplified two-UAV scenario. Suppose two UAVs are deployed at the same altitude $h$ with horizontal coordinates $\mathbf q_1 = (-d/2,0,h)$ and $\mathbf q_2 = (d/2,0,h)$, where $d$ denotes the horizontal inter-UAV distance between the two UAVs. To satisfy the safety-distance requirement, the feasible separation must satisfy $d \ge d_{\min}$.

\begin{figure}[!t]
    \centering
    \includegraphics[width=0.5\linewidth]{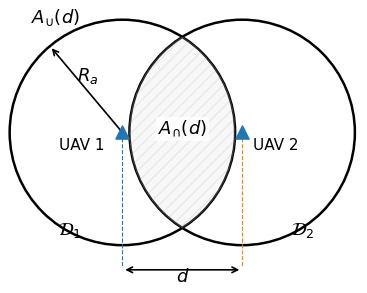} 
    \caption{Geometric illustration of the overlap and union areas of two UAV accessibility disks. Each disk has radius \(R_a\), and the two disk centers are separated by the horizontal inter-UAV distance \(d\). The shaded region denotes the overlap area \(A_{\cap}(d)\) in Eq.~(1), while the union of the two disks corresponds to \(A_{\cup}(d)\).}
\label{fig:two_uav_overlap}
\end{figure}

For analytical tractability, we approximate the accessible computing set of each UAV by an effective accessibility disk with radius $R_a(h)$ on the ground plane. Here, $R_a(h)$ denotes the maximum horizontal distance from the UAV projection within which a representative task can still satisfy the end-to-end latency constraint. This disk approximation is introduced only for geometric analysis of the two-UAV case and does not replace the exact accessibility definition in the general multi-UAV model. As in~\cite{Deng_2026}, we assume that ground CNs are distributed according to a homogeneous Poisson Point Process (PPP) with density $\lambda_c$, and the average computing capability of each CN is $\bar f$.
Let \(A_{\cap}(d)\) and \(A_{\cup}(d)\) denote the overlap
area and union area of the two accessibility disks, respectively.
As illustrated in Fig.~\ref{fig:two_uav_overlap}, the two disks have the
same radius \(R_a\), and their centers are separated by the horizontal
inter-UAV distance \(d\). The shaded lens-shaped region corresponds to
the duplicated accessible region \(A_{\cap}(d)\), while the union of the
two disks corresponds to \(A_{\cup}(d)\).
Then, for \(d_{\min}\le d < 2R_a\), the overlap area is
\begin{equation}
A_{\cap}(d)=
2R_a^2\cos^{-1}\!\left(\frac{d}{2R_a}\right)
-\frac{d}{2}\sqrt{4R_a^2-d^2},
\end{equation}
and the union area is
$A_{\cup}(d)=2\pi R_a^2-A_{\cap}(d)$.
For $d\ge 2R_a$, the two disks do not overlap and thus
$A_{\cup}(d)=2\pi R_a^2$.

\textbf{Lemma 1.}
Under the disk approximation, the expected unique accessible computing capacity is
$\mathbb E[\Psi(d)] \approx \bar f \lambda_c A_{\cup}(d)$.

\begin{proof}
Under the disk approximation, accessible CNs correspond to those located inside the union of the two accessibility disks. Since CNs follow a homogeneous PPP with density $\lambda_c$, the expected number of accessible CNs equals $\lambda_c A_{\cup}(d)$. Multiplying this quantity by the average computing capability $\bar f$ gives the result.
\end{proof}

\textbf{Theorem 1.}
For $d\in[d_{\min},2R_a)$, $\mathbb E[\Psi(d)]$ is strictly increasing and strictly concave in $d$.
\begin{proof}
From Lemma 1,
$\mathbb E[\Psi(d)] \approx \bar f \lambda_c A_{\cup}(d)$, $A_{\cup}(d)=2\pi R_a^2-A_{\cap}(d)$.
Differentiating $A_{\cup}(d)$ over $d\in[d_{\min},2R_a)$ yields
\begin{equation*}
\frac{\mathrm d \mathbb E[\Psi(d)]}{\mathrm d d}
=
\bar f \lambda_c \sqrt{4R_a^2-d^2}>0,
\end{equation*}
which implies that $\mathbb E[\Psi(d)]$ is strictly increasing. Differentiating again gives
\begin{equation*}
\frac{\mathrm d^2 \mathbb E[\Psi(d)]}{\mathrm d d^2}
=
-\bar f \lambda_c \frac{d}{\sqrt{4R_a^2-d^2}}<0,
\end{equation*}
which shows that $\mathbb E[\Psi(d)]$ is strictly concave. For $d\ge 2R_a$, the two disks do not overlap and $A_{\cup}(d)=2\pi R_a^2$ becomes constant. Hence, $\mathbb E[\Psi(d)]$ saturates for $d\ge 2R_a$.
\end{proof}

\textbf{Corollary 1.}
If the deployment objective is simplified to maximizing only the unique accessible computing capacity, while ignoring the communication-performance penalty, then the optimal UAV separation is $d^\star = \max(d_{\min},\, 2R_a)$.

Theorem 1 and Corollary 1 reveal that enlarging the UAV separation always increases the unique accessible computing capacity until the two accessibility disks no longer overlap. However, a larger separation also increases the average user--UAV distance, which tends to enlarge the uplink delay and may reduce the task completion probability. Therefore, when both computing accessibility and communication performance are considered, the optimal separation generally lies in the interval $[d_{\min},2R_a]$. This tradeoff provides the key motivation for the proposed cooperative computing accessibility-aware deployment design, which aims to enlarge the unique accessible computing set while suppressing excessive overlap and maintaining favorable communication conditions.

\section{CA3D: Computing-Accessibility-Aware Cooperative 3D Deployment}

The above analysis shows that cooperative UAV deployment should enlarge the unique accessible computing set while maintaining favorable service conditions for delay-constrained tasks. Excessive UAV overlap causes redundant accessibility to the same CNs, whereas overly dispersed deployment may degrade user access quality. Motivated by this tradeoff, we formulate a computing-accessibility-aware cooperative 3D deployment problem.

For a given deployment \(\mathbf Q\), we evaluate task completion using a task-level UAV-CN pair selection rule. For GU \(k\), let \(\mathcal A_k(\mathbf Q)\) denote the feasible UAV-CN pair set under the current deployment. Each feasible pair \((m,n)\in\mathcal A_k(\mathbf Q)\) has end-to-end latency \(D_{k,m,n}=D_{k,m}^{\rm up}+D_{m,n}^{\rm fwd}+D_{k,n}^{\rm cmp}\), and GU \(k\) selects
$
(m_k^\star,n_k^\star)
=
\arg\min_{(m,n)\in\mathcal A_k(\mathbf Q)}
D_{k,m,n}.
$
If \(\mathcal A_k(\mathbf Q)=\emptyset\), the task is unsuccessful; otherwise, the task success indicator is
\[
\mathbbm{1}_k(\mathbf Q)
=
\begin{cases}
1, & D_{k,m_k^\star,n_k^\star}\le D_k^{\max},\\
0, & \text{otherwise}.
\end{cases}
\]
Thus, \(P_{\rm succ}(\mathbf Q)=\frac{1}{K}\sum_{k=1}^{K}\mathbbm{1}_k(\mathbf Q)\). This rule is used to evaluate a given deployment rather than introducing the association as an additional optimization variable.

To reduce scale imbalance among different utility terms, we use a scale-calibrated deployment utility. The accessible computing capacity \(\Psi(\mathbf Q)\) and the overlap penalty \(\Omega(\mathbf Q)\) are evaluated in GHz, while \(P_{\rm succ}(\mathbf Q)\in[0,1]\) denotes the task success probability. For two UAVs \(m\) and \(m'\), the duplicated accessible computing capacity is $O_{m,m'}(\mathbf Q) = \sum_{n\in \mathcal C_m(\mathbf q_m)\cap \mathcal C_{m'}(\mathbf q_{m'})} f_n$.
The average pairwise overlap penalty is defined as
\[
\Omega(\mathbf Q)
=
\frac{2}{M(M-1)}
\sum_{m<m'}
O_{m,m'}(\mathbf Q),
\]
for \(M>1\), and \(\Omega(\mathbf Q)=0\) when \(M=1\). The cooperative 3D deployment problem is formulated as
\begin{equation}
\label{eq:opt_problem}
\begin{aligned}
\max_{\mathbf Q}\quad
& F(\mathbf Q)
= \alpha \Psi(\mathbf Q)
+ \beta P_{\rm succ}(\mathbf Q)
- \gamma \Omega(\mathbf Q) \\
\mathrm{s.t.}\quad
& h_{\min}\le h_m\le h_{\max},\quad \forall m\in\mathcal M,\\
& \|\mathbf q_m-\mathbf q_{m'}\|\ge d_{\min},\quad \forall m\ne m',
\end{aligned}
\end{equation}
where \(\alpha\), \(\beta\), and \(\gamma\) are scale-calibration and preference coefficients for accessible computing expansion, task completion, and overlap reduction, respectively.

The problem is highly non-convex because the accessible computing sets are discrete and deployment-dependent, while \(\Psi(\mathbf Q)\) and \(\Omega(\mathbf Q)\) couple the deployments of multiple UAVs. We therefore develop \emph{CA3D}, a two-stage heuristic algorithm combining global initialization and local refinement.

In the first stage, \textit{CA3D} performs PSO-based global initialization over the UAV deployment matrix \(\mathbf Q\). Each particle represents a candidate UAV layout and is updated using the standard PSO rule with inertia weight \(\omega\), cognitive/social coefficients \(c_1,c_2\), and projection \(\Pi_{\mathcal Q}(\cdot)\) onto the feasible deployment set. The best particle after global search is used to initialize the second stage.
In the second stage, \textit{CA3D} refines the deployment by beam-search-based local updates. For UAV \(m\), the local action set is
$\mathcal{A}_m
=
\{(\pm\Delta x,0,0),(0,\pm\Delta y,0),(0,0,\pm\Delta h),(0,0,0)\}$.
For a candidate action sequence \(\mathbf a_{m,1:H}\), the discounted rollout score is
$J_m(\mathbf a_{m,1:H})
=
\sum_{\ell=1}^{H}
\rho^{\ell-1}
\big(F(\mathbf Q^{(\ell)})-F(\mathbf Q)\big)$,
where \(H\) is the rollout horizon, \(\rho\) is the discount factor, and \(\mathbf Q^{(\ell)}\) is the rollout deployment. Only the best \(W\) partial sequences are retained at each depth, and the first action of the highest-scoring sequence is executed. This process is repeated for all UAVs until convergence.

A concise complexity analysis is as follows. The complexity of one deployment-utility evaluation is \(C_F=O(KMN+M^2N)\), where the two terms correspond to GU-UAV-CN latency checking and pairwise overlap evaluation, respectively. Therefore, the overall complexity of CA3D is \(O((N_pI_g+I_bMHW A_{\rm loc})C_F)\), where \(N_p\), \(I_g\), \(I_b\), \(H\), \(W\), and \(A_{\rm loc}\) denote the number of particles, global-search iterations, beam-search refinement iterations, rollout horizon, beam width, and local-action size, respectively.

\section{Performance Evaluation}
\begin{figure}[!t]
    \centering
    \begin{minipage}{\linewidth}
        \centering
        \includegraphics[width=0.725\linewidth]{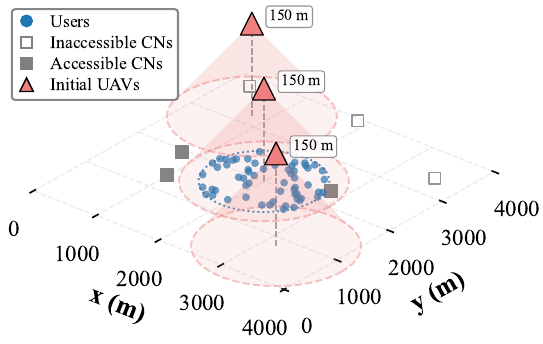}
        \caption{Initial random deployment with three UAVs.}
        \label{fig:3d_initiation}
    \end{minipage}

    \vspace{1em} 

    \begin{minipage}{\linewidth}
        \centering
        \includegraphics[width=0.725\linewidth]{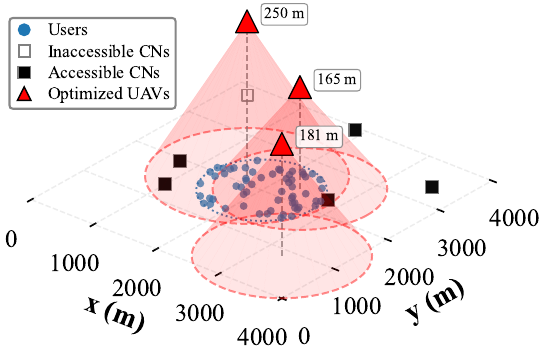}
        \caption{CA3D-optimized deployment with three UAVs.}
        \label{fig:3d_optimized}
    \end{minipage}
\end{figure}

\begin{figure*}[tb]
    \centering

    \begin{subfigure}[b]{0.5\textwidth}
        \centering
        \begin{subfigure}[b]{0.5\textwidth}
            \centering
            \includegraphics[width=\linewidth]{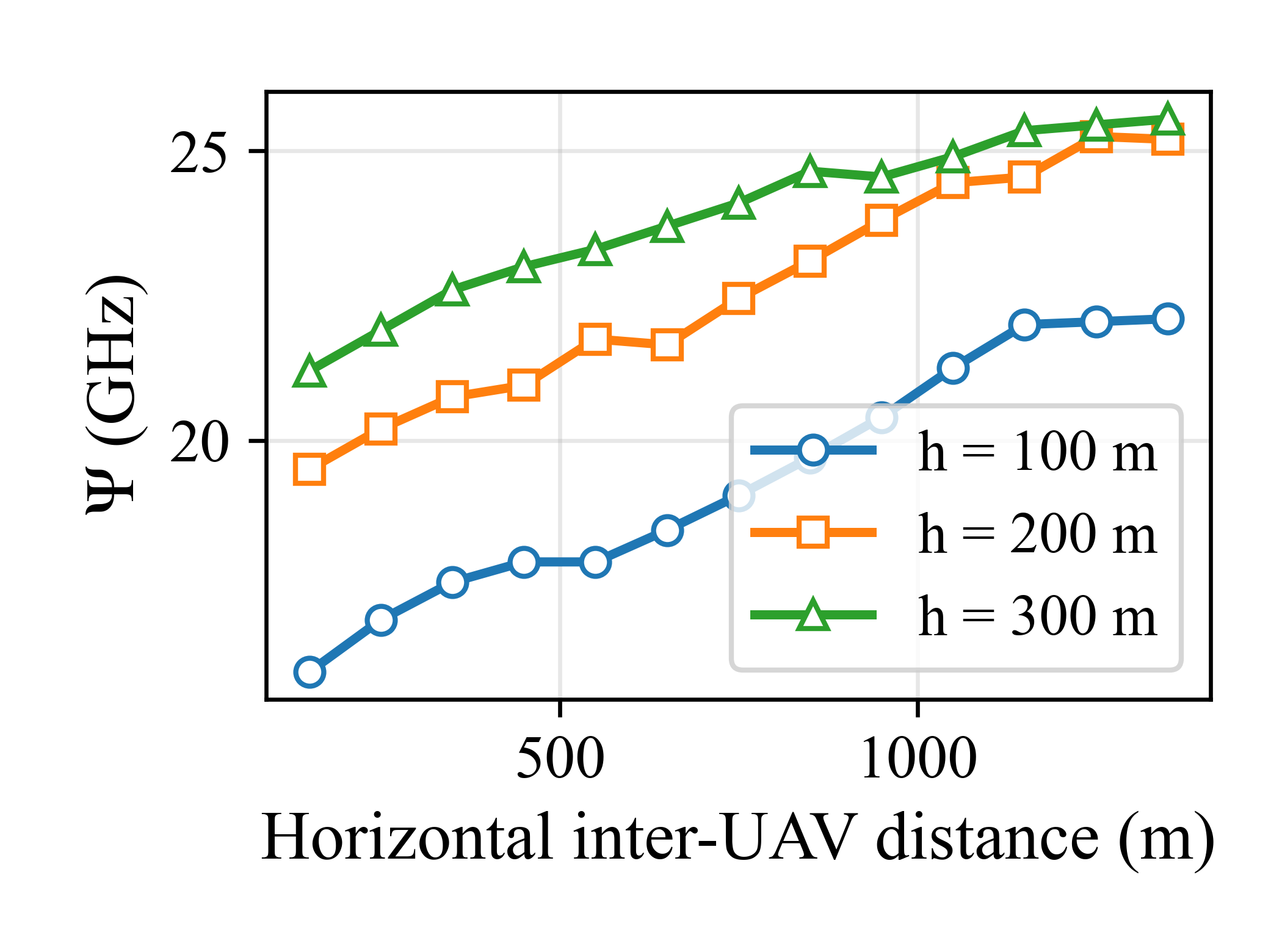}
            \subcaption{Accessible computing capacity.} 
            \label{fig:4a}
        \end{subfigure}
        \hspace{-4mm} 
        \begin{subfigure}[b]{0.5\textwidth}
            \centering
            \includegraphics[width=\linewidth]{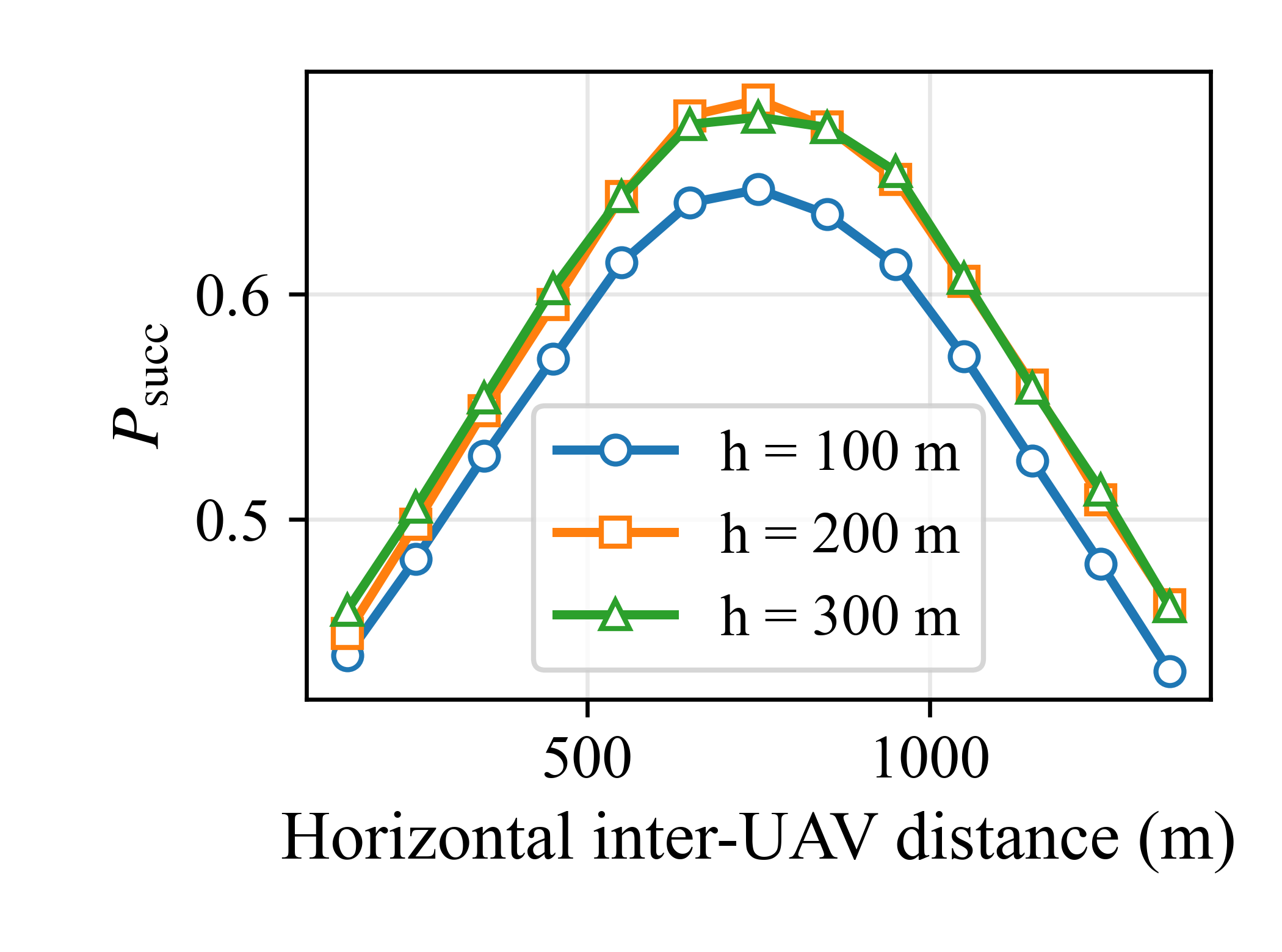}
            \subcaption{Task completion probability.} 
            \label{fig:4b}
        \end{subfigure}

        \caption*{Fig. 4: Impact of horizontal spacing between two UAVs on performance.}
    \end{subfigure}
    \hspace{-4mm} 
      \setcounter{subfigure}{0} 
    \begin{subfigure}[b]{0.5\textwidth}
        \centering
        \begin{subfigure}[b]{0.5\textwidth}
            \centering
            \includegraphics[width=\linewidth]{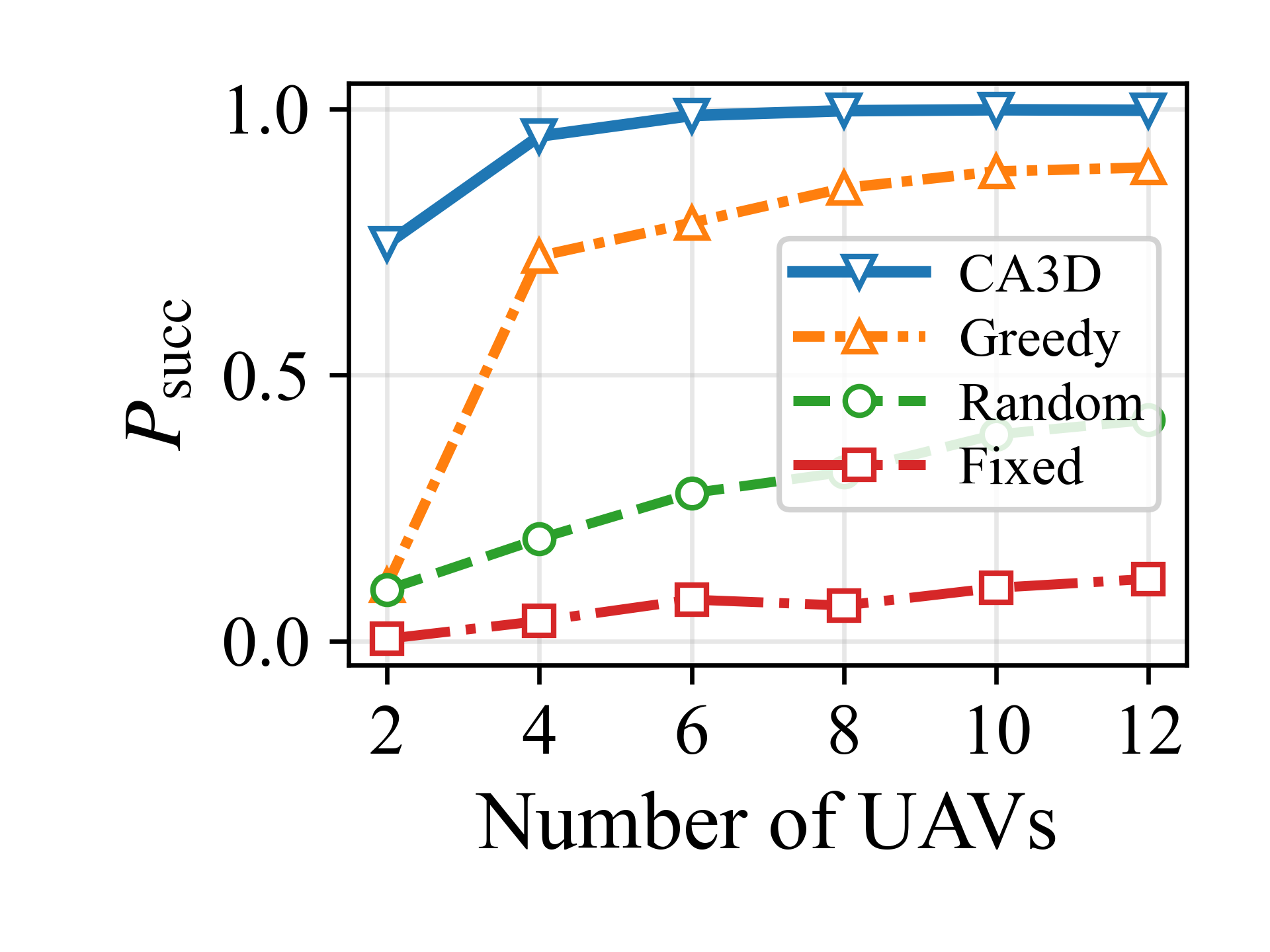}
            \subcaption{Hotspot GU distribution.} 
            \label{fig:4a}
        \end{subfigure}
        \hspace{-4.8mm} 
        \begin{subfigure}[b]{0.5\textwidth}
            \centering
            \includegraphics[width=\linewidth]{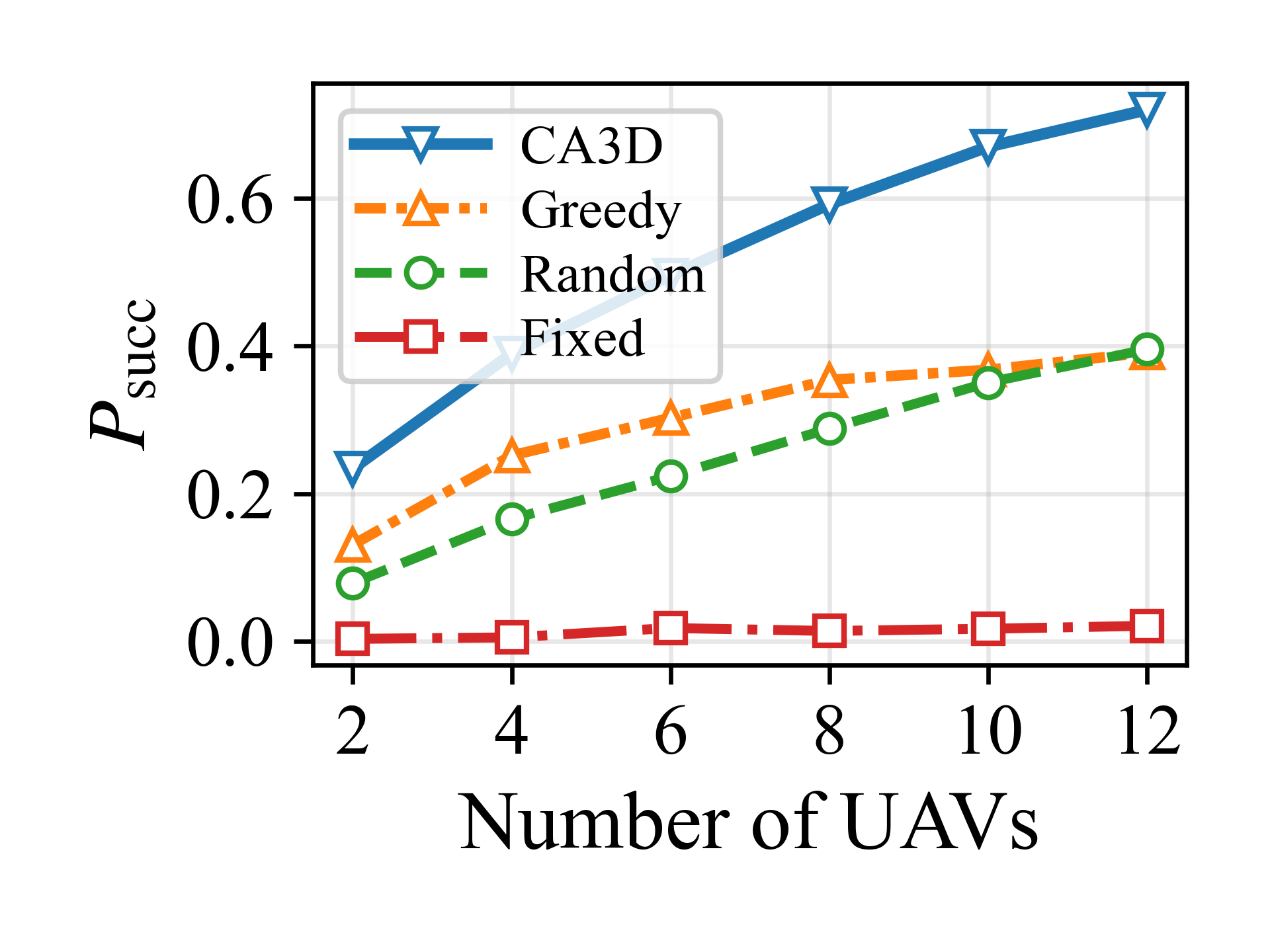}
            \subcaption{Random GU distribution.} 
            \label{fig:4b}
        \end{subfigure}

        \caption*{Fig. 5: Impact of GU distribution on task completion vs. number of UAVs.}
    \end{subfigure}
    \label{fig:Fig4_overall}
\end{figure*}

In this section, we evaluate the performance of the proposed computing-accessibility-aware cooperative UAV deployment framework and verify the theoretical insights. 
We consider a \(4\,\mathrm{km}\times4\,\mathrm{km}\) district-level service area, such as a campus, traffic hub, industrial park, or emergency service region, where local task demand may need to be connected to geographically distributed CNs~\cite{Sun2024joint}.
Unless otherwise stated, GUs are uniformly distributed within a centered hotspot disk of radius \(800\) m, while CNs are randomly distributed over the entire area. 
This hotspot setting models locally concentrated task requests with insufficient or congested nearby terrestrial computing resources, which motivates the use of UAVs to expand the accessible computing resource pool. 
The UAV altitude is constrained within \([100,300]\)~m, and the bandwidth is \(10\)~MHz. 
Each task has an input size of \(5\)~MB, a computation requirement of \(1\times10^9\) CPU cycles, and a deadline of \(1\)~s. 
The available computing capability of each CN is independently generated as \(f_n\sim\mathcal U(2,10)~\mathrm{GHz}\).
For the scale-calibrated utility in Eq.~\eqref{eq:opt_problem}, we set \(\alpha=0.2\), \(\beta=1.5\times10^4\), and \(\gamma=0.1\). Here, \(\Psi(\mathbf Q)\) and \(\Omega(\mathbf Q)\) are evaluated in GHz, and \(P_{\rm succ}(\mathbf Q)\in[0,1]\).

Fig.~\ref{fig:3d_initiation} and Fig.~\ref{fig:3d_optimized} show a representative random initial deployment and the corresponding \textit{CA3D}-optimized deployment with three UAVs. Compared with the initial layout, \textit{CA3D} relocates UAVs to better connect the hotspot GU region with distributed CNs, making more remote CNs accessible and reducing redundant service overlap.

Fig.~4(a) shows that the accessible computing capability \(\Psi\) increases with the horizontal inter-UAV distance and gradually saturates as the overlap between the two accessibility disks decreases, which is consistent with Theorem~1. A larger altitude generally leads to a larger \(\Psi\) because it enlarges the effective accessibility radius. In contrast, Fig.~4(b) shows that \(P_{\rm succ}\) is non-monotonic with respect to UAV spacing. It first improves as two UAVs become more complementary in serving users and accessing distributed CNs, but then decreases when the spacing becomes too large due to degraded user--UAV link access quality. The maximum task completion probability is achieved at a moderate spacing of around \(700\)~m, confirming the tradeoff between expanding accessible computing resources and maintaining favorable service conditions.

Fig.~5 compares the task success probability \(P_{\rm succ}\) versus the number of UAVs under hotspot and random GU distributions for \textit{CA3D} and three baselines: \textit{Random}~\cite{Han2024}, \textit{Fixed}~\cite{lu2024resource}, and \textit{Greedy}. Specifically, \textit{Random} randomly deploys multiple UAVs without cooperative optimization; \textit{Fixed} restricts the candidate accessible CNs to a fixed local region with radius \(200\) m; and \textit{Greedy} uses the same local action set as \textit{CA3D}, but removes the PSO-based global initialization and multi-step beam-search rollout, and greedily selects the one-step movement with the largest immediate improvement in \(F(\mathbf Q)\).
Under the hotspot distribution in Fig.~5(a), \textit{CA3D} rapidly approaches nearly full task completion as \(M\) increases. When \(M=8\), \textit{CA3D} achieves \(P_{\rm succ}\approx 1.0\), compared with about \(0.83\), \(0.30\), and \(0.04\) for \textit{Greedy}, \textit{Random}, and \textit{Fixed}, respectively. Under the random-GU distribution in Fig.~5(b), the enlarged service region makes the problem more challenging, but \textit{CA3D} still achieves the best performance. When \(M=12\), \textit{CA3D} reaches \(P_{\rm succ}\approx 0.70\), while \textit{Greedy} and \textit{Random} stay around \(0.35\)--\(0.40\), and \textit{Fixed} remains close to zero. The consistent gains over all three baselines under both GU distributions demonstrate the robustness of \textit{CA3D}, which jointly improves task success, expands accessible computing capacity, and reduces redundant UAV overlap.

\section{Conclusion}
This letter has studied computing-accessibility-aware cooperative 3D deployment of multiple UAVs for ground computing systems. We have shown that UAV deployment reshapes both communication quality and the reachability to distributed CNs under latency constraints. The proposed \textit{CA3D} design has jointly balanced accessible computing capacity, task completion probability, and redundant UAV overlap. Simulation results under heterogeneous CN capacities and different GU distributions have confirmed that \textit{CA3D} consistently outperforms representative baselines, demonstrating the value of computing accessibility as a design principle for multi-UAV ground computing systems.


\bibliography{deng}

\end{document}